\documentclass[a4paper,twoside]{article}

\usepackage{epsfig}
\usepackage{subfigure}
\usepackage{calc}
\usepackage{amssymb}
\usepackage{amstext}
\usepackage{amsmath}
\usepackage{amsthm}
\usepackage{multicol}
\usepackage{pslatex}
\usepackage{apalike}
\usepackage{url}
\usepackage{SCITEPRESS}
\usepackage[small]{caption}

\begin{document}

\title{Cyber-Virtual Systems: Simulation, Validation \& Visualization}

\author{\authorname{Jan Olaf Blech, Maria Spichkova, Ian Peake, Heinz Schmidt}
\affiliation{RMIT University}
\email{\{janolaf.blech, maria.spichkova, ian.peake, heinz.schmidt\}@rmit.edu.au}
}


\abstract{We describe our ongoing work and view on simulation, validation and visualization of cyber-physical systems in industrial automation during development, operation and maintenance. System models may represent an existing physical part -- for example an existing robot installation -- and a software simulated part -- for example a possible future extension. We call such systems cyber-virtual systems. 
In this paper, we present the existing VITELab infrastructure for visualization tasks in industrial automation. The new methodology for simulation and validation motivated in this paper integrates this infrastructure. We are targeting scenarios, where industrial sites which may be in remote locations are modeled and visualized from different sites anywhere in the world. 
Complementing the visualization work, here, we are also concentrating on software modeling challenges related to cyber-virtual systems and simulation, testing, validation and verification techniques for them. 
Software models of industrial sites require behavioural models of the components of the industrial sites such as models for tools, robots, workpieces and other machinery as well as communication and sensor facilities. Furthermore, collaboration between sites is an important goal of our work.
}

\onecolumn \maketitle \normalsize \vfill

\section{\uppercase{Introduction}}
\label{sec:introduction}

Operation, development, maintenance (including modifications and extensions) of industrial automation facilities like factories or mining sites profit from software support such as software based monitoring, controlling and collaboration tools. This requires visualization capacities as well as software models of the physical entities involved and ways to reason about them.
Industrial automation facilities typically comprise machinery like robots and their components. Components may serve as actuators: tools, conveyor belts, work pieces or pipes, valves and pumps in cases were fluids or gases are processed.  Sensors can be found throughout industrial automation sites. The data gathered from the sensors may be stored in a central facility.

Hardware-in-the-loop (HIL) approaches \cite{schlager2008hardware} are now standard in the development of system components in domains such as automative systems, e.g.,  \cite{isermann1999hardware}, avionics and also in industrial automation. In HIL, parts of a system are simulated in software to test a distinct system component. 
In this paper, we are going one step further and aim at simulating different parts of an industrial site. 
We do not restrict our approach to the development, but also aim at supporting operation and maintenance of industrial automation facilities. Furthermore, we aim at visualizing remote facilities or parts of them. This is especially crucial when developing, operating or maintaining industrial sites located in areas that are difficult to access such as mines and oil rigs and for collaboration between different sites and sharing knowledge between them. 

In the case  where components of a system are manufactured at different places,
transport from component development and production locations to integration and deployment sites can significantly increase the whole development costs as well as time. Integration can reveal additional work tasks and further  transportation of the system's parts may be necessary.
If a system's components are bulky or heavy, this may also delay optimization and correction. 

For this reason, we present an existing visualization infrastructure -
the Virtual Interoperability Test Lab (VITELab)\footnote{VITELab is an eResearch facility of the Australia-India Research Centre for Automation Software Engineering (AICAUSE), a partnership between RMIT University and the ABB Group (Australia and India), \url{http://rmit.edu.au/research/aicause}.} a global laboratory
connecting industry and university sites and providing a collaboration
platform for experimental design and testing of cyber-physical
systems. Among its aims are to reduce development costs by simulating
and virtually testing possible deployments before the system is
actually physically set up.
We also present the corresponding new and ongoing research directions towards combining visualization and software support for reasoning about industrial automation facilities. 
The ideas featured in this paper comprise the following ingredients:
\begin{itemize}
\item The use of VITELab, in particular the Global Operations
  Visualization (GOV) Lab, a high resolution multi-screen
  visualization facility. 
\item Software models for system components that comprise spatio-temporal information about a component's behavior and ways to reason about them, testing and simulation.
\item The combination and integration of these for industrial automation.
\end{itemize}
Our work is a step towards software solutions facilitating global collaboration between developers, operators and maintenance of industrial sites.

\section{\uppercase{Related Work}}

\textbf{Modelling aspects} Different languages exist for the modeling of embedded and automation systems. Standards like IEC 61131-3 and IEC 61499 target the software part of control systems and thus specify the behavior of machinery.
In the scientific community different modeling languages such as the Petri-Net semantics based BIP \cite{basu2006modeling} for distributed asynchronous systems and 
Modelica, providing means for modeling and simulation of systems have been established, 
cf. \cite{modelica2008}, \cite{Fritzson2004Modelica}, \cite{modelica2013vehicle}.  
Modelica is object-oriented
and its latest extensions allow  modelling of system requirements~\cite{modelica2013req} as well as 
simulation of technical and physical systems~\cite{Fritzson2011Modelica}.
Modeling theories for distributed hybrid system such as SHIFT~\cite{Deshpande97shift:a} and R-Charon~\cite{KratzSPL06} guarantee a complete simulation and compilation of the models,
but do not support verification or analysis of the system on the modeling level.
Same limitations also apply to the input language of the model checkers  UPPAAL~\cite{DBLP:conf/sfm/BehrmannDL04}  and PHAVer~\cite{Beek06syntaxand}: the verification capabilities do not match the whole expressiveness of the modeling languages.

Assigning semantics to logical entities for categorizing and reasoning about them is a one goal of our models for industrial automation facilities. The concept has been made popular in the context of the semantic web \cite{berners2001semantic} and ontologies \cite{staab2001knowledge}.

\textbf{Spatial aspects} 
The modeling of industrial automation sites involves spatial aspects. For example, robots must ensure a behavior that guarantees collision avoidance and the correct handling of workpieces. Systems that comprising thermal aspects like heat exchangers need adequate models to cover their behavior.
SpaceEx \cite{frehse2011} allows
the modeling of continuos hybrid systems based on hybrid automata. It
can be used for computing overapproximations of the space occupied by objects.
A process algebra for 3D objects is provided in
\cite{cardelli2010processes}. 
Results on spatial interpretations are explained in
\cite{hirschkoff}. A quantifier-free rational fragment of  logic suitable for describing spatial scenarios has been shown to be decidable in \cite{zilio}.
Logics for spatio-temporal reasoning go back to the
seventies. The Region Connection Calculus (RCC) \cite{bennett}  includes
spatial predicates of separation. RCC features predicates
indicating that regions do not share points at all, points on the
boundary of regions are shared, internal contact  where one region is included
and touches on the boundary of another from the inside,  overlap of
regions,
and  inclusion.

\textbf{Cyber-physical aspects} 
Many approaches on mechatronic/cyber-physical systems omit an  abstract logical level of the system representation and lose the advantages of the abstract representation. The work presented in~\cite{Vogel-Heuser_IECON} defines an extensive support to the components communication and time requirements, while the model discussed in~\cite{IEEE_INDIN_2011} proposes a complete model of the processes with communication. 
In traditional development of embedded systems e.g., \cite{ES_Berger}, 
the system is usually separated into software and hardware parts as soon as possible, at an early stage of the development process. This does not always benefit the development process, because when using an abstract level of modeling  the difference in the nature of components does not necessarily play an important role. 
\cite{Sapienza2690} and \cite{Spichkova_Campetelli2012} independently suggest to use a platform-independent design in the early stages of system development. 
The approach presented in \cite{Sapienza2690} introduces the idea of pushing
hardware- and software-dependent design  as late as possible, 
however, the question of the current practical and fundamental limitations of logical modeling in comparison to cyber-physical testing, 
is not completely answered. 
In comparison to \cite{Sapienza2690}, 
the focus of \cite{Spichkova_Campetelli2012}  
is on reutilisation and generalisation of two existing software systems development methodologies 
(both elaborated according to the results of the case studies motivated and supported 
by DENSO Corporation and Robert Bosch GmbH)  
for application within the cyber-physical
domain to benefit from the advantages these techniques have shown. 
The question, how deep we can go on the modeling of cyber-physical systems on the logical level is still open in both approaches.
The goals presented here are also related to  hybrid commissioning \cite{dominka2007hybrid}.

\textbf{Early analysis aspects} 
The idea of early analysis of critical system faults has the goal to identify faults which mutate the safety critical behaviour of the system, 
and to identify test scenarios which can expose such faults from an abstract modeling level,
i.e. by generation of tests (both for real system and its model) from formal specifications or from the CASE tool models (cf., e.g., \cite{Daggupta2012tests,Broy2005testing,pretschner200510}). 
The approach has certain limitations due the abstract nature of the formal model serving as a base for the test generation as well as 
an underlying assumption of existence of a precise formal model of the system being developed. Even when taking into account these limitations and assumptions, 
these approaches allow automatization of test case design and make the design process more stringent. 
VITELab and the described research complements commercially available visualization software for collaboration purposes in industrial automation such as DELMIA\footnote{\url{http://www.3ds.com/products-services/delmia/products/all-delmia-products/}}. The approach described here, is building on (semi-)formal models which carry semantic meaning and are suitable for automatic interpretation and processing, whereas the DELMIA focus is even more on visualization.

\section{\uppercase{From cyber-physical to cyber-virtual Systems}}

\begin{figure}
 \centering
\includegraphics[width=0.46\textwidth]{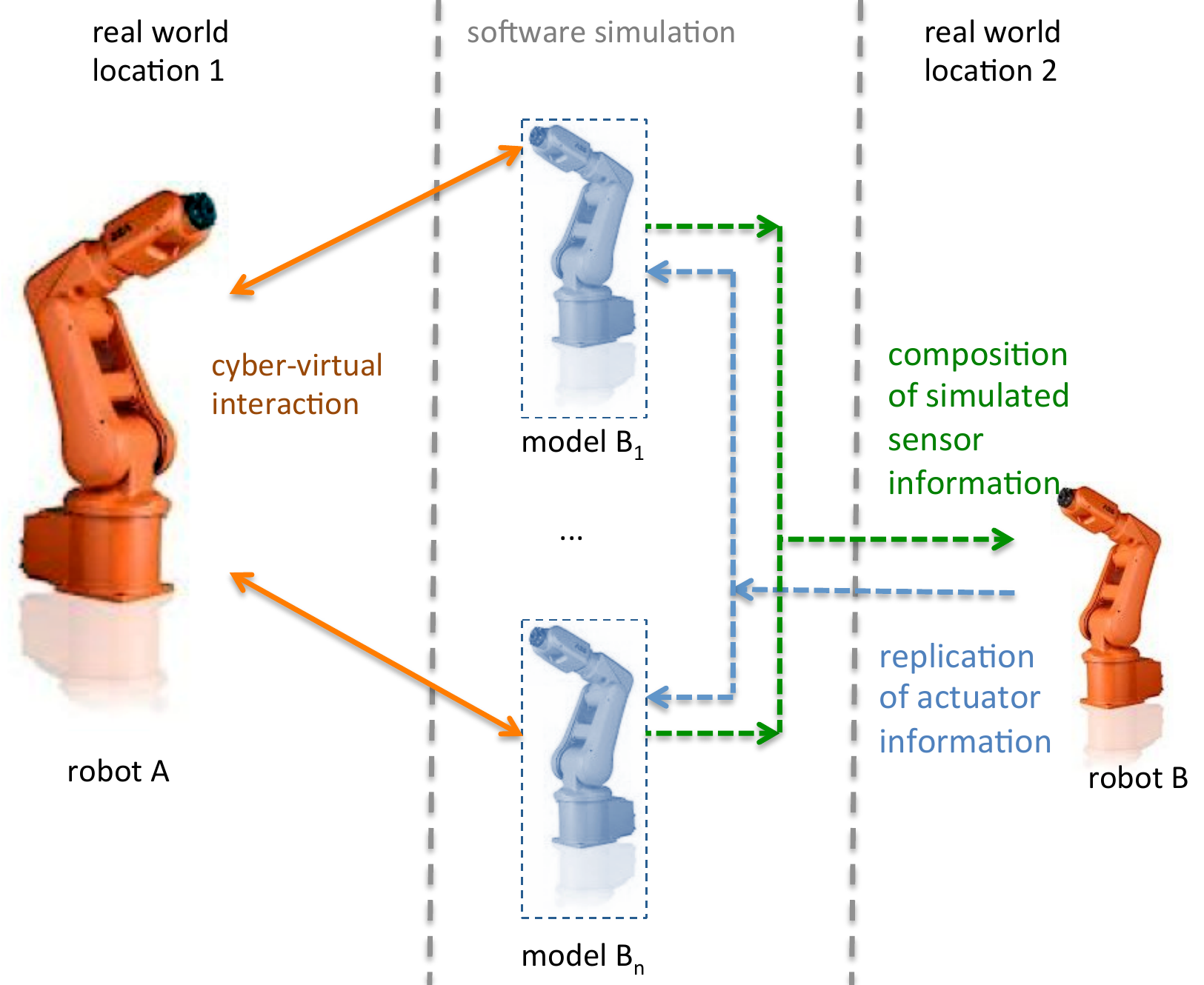}
 \caption{Cyber-virtual communication}
 \label{fig:example}
\end{figure}

Let us discuss an example scenario based on the ideas of the virtual interoperation testing.
In an industrial plant we require the integration/interoperability of $n+1$ bulky/heavy robots (cf. Figure~\ref{fig:example}): 
a robot of the type $AType$ (lets call it robot $A$) is assembled in location $L_{A}$, 
the $n$ other robots are of a different type $BType$ and are assembled in a different location location $L_{B}$.
The robots are in different locations and making them work together in a different shared deployment location requires extensive simulation, testing and collaboration.

Assuming in addition that the $n$ robots of type $BType$ perform simultaneously similar movements and actions (e.g., they stamp similar details on workpieces on a conveyor belt and are doing the same movements, even in the case their stamps are different), 
we can simulate their behaviour using a single robot $B$: its actuator information will be replicated to obtain 
$n$ virtual models $B_{1}, \dots, B_{n}$, and its sensor information will be extended by the composition of the modeled sensor information from $B_{1}, \dots, B_{n}$. The sensor information of the robot $A$ will be a composition of the real sensor data and the sensor data modeled according to the actions of $B_{1}, \dots, B_{n}$.

Thus, to check the interoperability of the robot $A$ and $n$ robots of the type $BType$ on the level of virtual interoperability testing, we need only two real robots: a robot $A$ and a robot $B$. Moreover, they could be located in $L_{A}$ and $L_{B}$ respectively, because the simulator and visualization facility may take the role of a physical medium between them, allowing to ignore the real distance between robots and also allowing to have a visualisation of the test and simulation not only at $L_{A}$ and $L_{B}$, but also on the third place $L_{C}$, where the corresponding laboratory is located. 

\begin{figure} 
\centering
\includegraphics[width=0.465\textwidth]{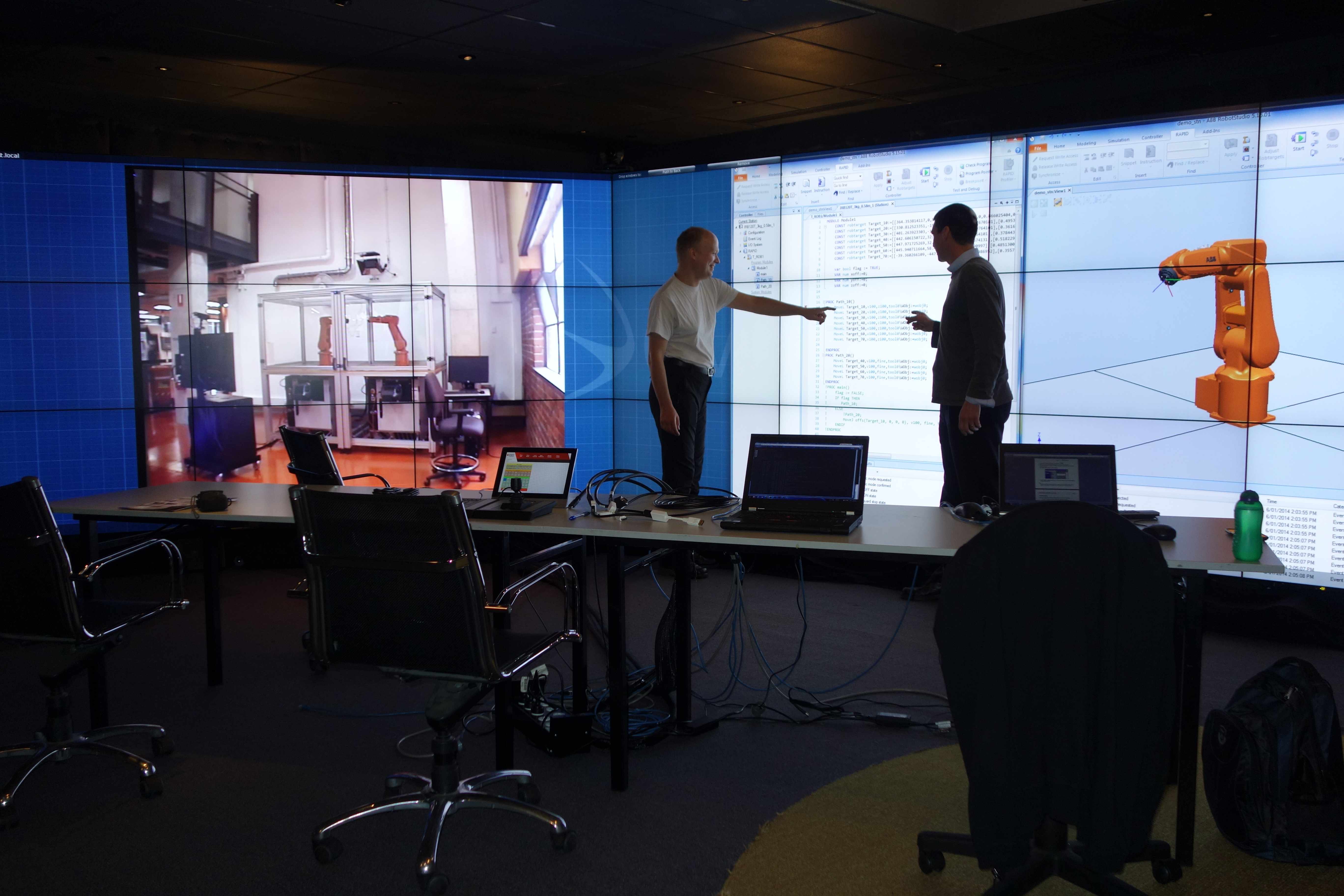}
\caption{VITElab in operation}
\label{fig:vitelab2}
\end{figure}

General ideas for using the virtual interoperability test lab (VITELab)  for the use of remote cyber-physical
integration/interoperability testing in a virtual environment 
 as a middle step between an abstract modelling and real testing were presented in \cite{issec2013spichkova}.
Figure~\ref{fig:vitelab2} 
shows the VITELab facility in operation, viewed from the GOV Lab.
VITELab gives a platform for a new level of simulation and integration:
interoperability simulation and testing is performed early and remotely,
for example while cyber-physical components are in the prototyping stage i.e. on the workbench: 
individual components (e.g., robots, manufacturing cells),
are connected in a suitable virtual environment, without being deployed at the same place physically.
Successful testing and simulation could significantly reduce the well-documented costs arising from discovery of design faults after implementation.

Research connected to VITElab is influenced by larger
cooperations in the industrial automation domain. 
Remote integration and testing allows for an integration and testing phase of a real system  assuming a certain level of abstraction where
the network, the virtual environment and the remote embodiments may be abstractions themselves. This level of abstraction includes real physical components of the system (in the case of the VITElab project, e.g., real robots and production plants) and more characteristics of the network, environment and embodiments. 
Our models and their visualization can give us the possibility to identify
{\it (i)} a number of problems and inconsistencies on the early stage of system development and verify especially important system's properties before the real system is build and integrated, and
{\it (ii)} possible weak points in the system (such as some timing properties, feature interactions, component dependancies) which we should focus on, during the testing phase. 
%
%

\section{\uppercase{Research Challenges and Corresponding Projects}}
This section presents research challenges connected to cyber-virtual
systems, VITELab, simulation and validation  in more detail. 
\paragraph{Main directions for research}
We have identified the following research challenges in our scenario:
\begin{itemize}
\item Simulation and the visualization of simulation runs.
\item Testing, verification and validation of cyber-virtual scenarios.
\item Gaining expertise and knowledge from joint work using visualization and simulation. 
\item Sharing and making expertise and knowledge available for similar development projects and for related operation and maintenance tasks in related facilities.
\end{itemize}
\paragraph{Software Models for Industrial Plants}
In our work, we propose two ingredients related to software models for addressing these challenges:
\begin{itemize}
\item  (Semi-)formal descriptions based on human factors approaches to achieve better readability/usability and understandability.  
\item Spatial behavioral models that capture the characteristics of entities and components in industrial automation. We are interested in establishing a type system for these components.
\end{itemize}
\paragraph{Existing VITELab projects}
The research challenges identified in the context of VITELab fall into the network, cloud and distributed computing areas, and are covered by the following ongoing projects:
\begin{itemize}
\item
{\it Network connectivity} between sites with specialist equipment is supported by dedicated links and research software stacks. 
\item
The {\it Cyber-physical Simulation Rack (CSRack)}, is a multi-node cloud server rack with attached RAID storage provides parallel cloud computing capability to support modeling and simulation and the capability to act as a 'cloudlet' gateway to major national and international cloud facilities such as NeCTAR\footnote{National eResearch Collaboration Tools and Resources Project, \url{https://www.nectar.org.au}}.
\item
The {\it Global Operations Visualization (GOV) Lab} project, provides videoconference and streaming capability to remote sites combined with a large high resolution tiled display wall.
\item
The  {\it Advanced Manufacturing Robot Interoperation Test (AMRIT) lab} provides industrial robots connected to the GOV lab. The robots comprise arms, sensors and cameras as ``eyes on the robots''.
 \end{itemize}
Further research challenges exist in the connection of software based development tools for industrial automation systems to the described infrastructure. Such tools may need to undergo a redesign of the software architecture to enable this, cf.  \cite{peaketowards}.

\section{From (Semi-)Formal Methods to Visualization \& Validation}
\begin{figure} 
 \centering
\includegraphics[width=0.35\textwidth]{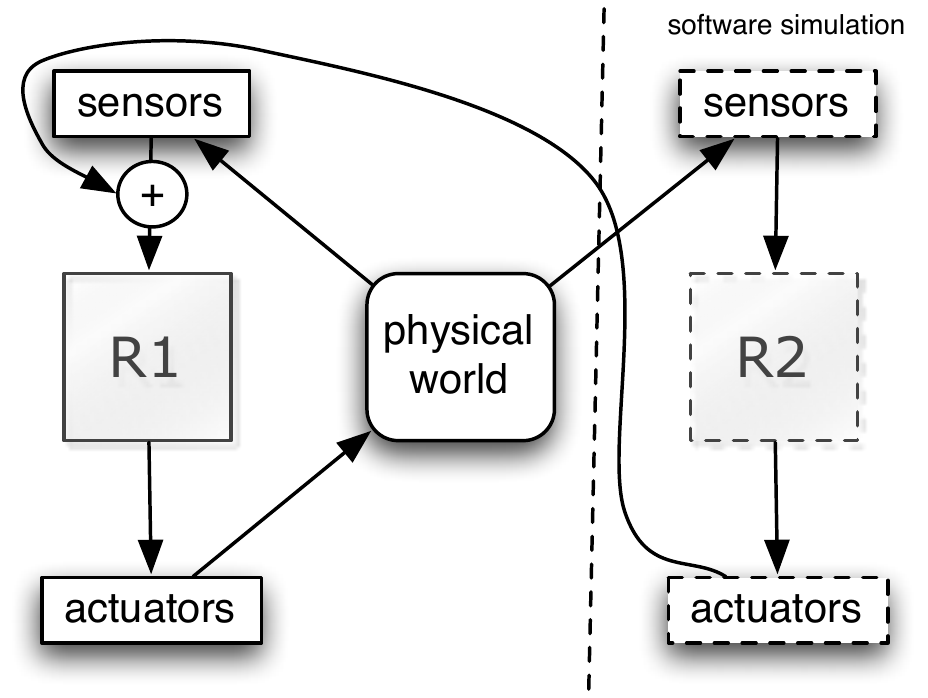}
 \caption{Robot in the loop}
 \label{fig:rlo}
\end{figure}

A starting point for our work is a HIL approach and is depicted in Figure~\ref{fig:rlo}. Here, the interplay of a physical robot with a virtual simulated robot is shown. The actions of the physical robot to the environment are observed passed to the robot simulation and reacting actions are calculated. These actions are (by)passed to the sensors of the physical robot to simulate the interplay.
The interplay can be analyzed both by software tools as well as human inspection.
The human based analysis profits from visualisation capabilities for the display of the simulated robot and the monitoring of the physical counterpart.

\paragraph{Human Factors and Formal Models}
To enable simulations we need (semi-)formal descriptions of robot behavior, which should not only fit for the simulation purposes but also be readable for system/verification engineers. In our approach we follow the ideas based on human factor analysis within formal methods \cite{Spichkova2013HFFM,hffm_spichkova}. This
allows to have short and  readable specifications of component behavior. It
 is appropriate for  switching between different modeling, specification and programming languages and
 is suitable for the application of specification, reasoning and proof methodology  \cite{spichkova2013we,spichkova}. 

\paragraph{Formal Proofs and Verification}
In the case of formal proofs, one of the main points of this methodology is an alignment of the future proofs during the specification phase
to make the proofs simpler and appropriate for application in practice. 
One direction for reasoning about a system represented in a formal specification framework, 
is the verification of its properties by translating the specification to a Higher-Order Logic and 
subsequently using the theorem prover following \cite{IsabelleAFP2013}. 

\begin{figure} 
 \centering
\includegraphics[width=0.431\textwidth]{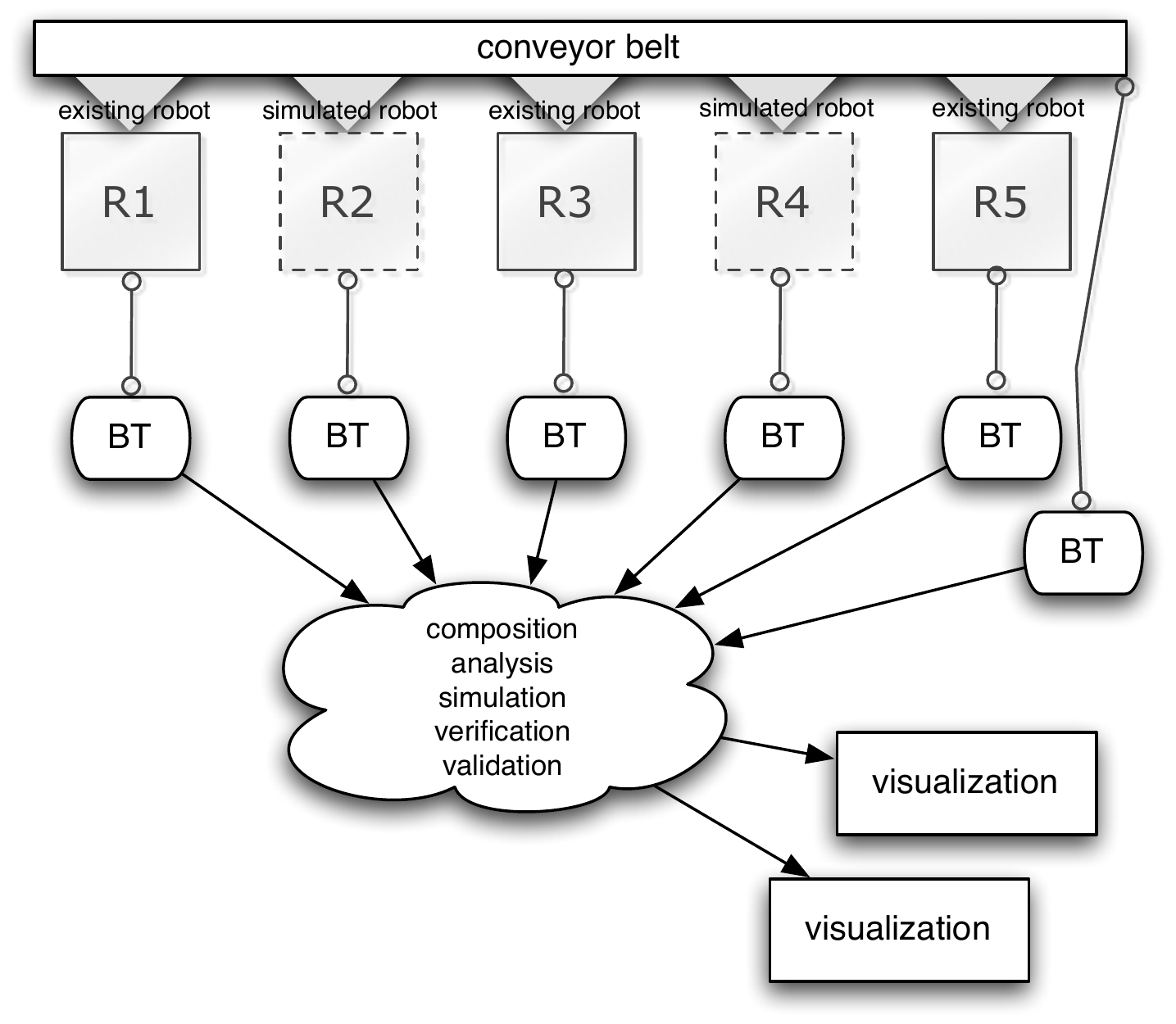}
 \caption{Combining virtual and physical robots with BT}
 \label{fig:scenario1}
\end{figure}

\paragraph{Spatial Behavioral Types}
Our (semi-)formal models comprise spatial behavioural.  This can be assigned to both physical and virtual simulated robots, their components and other entities  interacting with them as shown in Figure~\ref{fig:scenario1}. 
Following the ideas presented in \cite{blech2012behavioral} these spatial behavioural models can serve as a type system similar to types systems in higher programming languages like C and Java which come with basic types like integers, Strings and floating point values as well as composed types like records or classes. Here, we regard (spatial) Behavioural Types (BT). BT act as types for virtual or physical entities in our automation scenarios. They are characterised by the following core concepts:
\begin{itemize}
\item {\it Abstraction}. BT represent aspects of robots, robot components and other entities in industrial automation. BT abstract from details concerning interactions and internal structure.
\item {\it Conformance}. Type conformance of BT is used to relate entities in industrial automation correctly to a BT.
\item {\it Refinement}. BT should comprise a notion of spatio-behavioral refinement that allows replacing a component by a refined one. For example, the concept of refinement shall allow replacing a robot by a newer version that essentially provides the same functionality plus some new features.
\item {\it Compatibility}. Compatibility checking of BT is used to decide whether a component does indeed match required needs based on provided and expected BT. It should be decidable and automatic. 
\item {\it Inference}. A BT framework should allow to infer composed BT. For example, the BT of a robot may be inferred from the BT of its components.
\end{itemize}
\paragraph{Spatial Behavioural Types for Simulation and Validation}
BT  can serve as a specification basis for the components of robots and the robots composed of them. 
BT can be used to build models of industrial automation facilities. Using BT based specifications, we can perform:
\begin{itemize}
\item Simulation and {\it visualization} for human inspection and collaboration between developers, operators and maintenance personnel.
\item Automatic spatio-temporal reasoning for {\it collision detection} of robots and other entities.
\item Checking automatically the required {\it sensor ranges} and regions affected by physical entities.
\item Guaranteeing correct {\it interplay of tools and workpieces} in time and space.
\item Simulating the replacement of an entity such as a robot arm by another (refined) version.
\item Documenting behavior of system installations and sharing this for collaboration.
\end{itemize}
The BT concept is following the idea of interface automata
\cite{deAlfaro:2001:IA:503271.503226}.
It has been proposed as a type system for OSGi systems in the past 
\cite{blech2012behavioral}.
Theorem prover export and interactive verification of properties were
studied  in \cite{blech2012towards} and may be an issue for future work together with human-factor analysis.
Checking compatibility and means to make behavioral system
descriptions compatible were examined  in
\cite{DBLP:journals/corr/abs-1302-5175}. 
For checking the spatio-temporal properties in our scenarios we
incorporate the BeSpaceD \cite{blechtowards} tool. 
Checks in BeSpaceD are done by converting spatio-temporal models or BT and required properties into SMT and SAT problems and applying suitable solving techniques such as the z3 SMT solver \cite{de2008z3}.

\section{\uppercase{Conclusions}}
\label{sec:conclusion}
The presented research is ongoing work and part of larger cooperations
with an industrial automation company.
In this paper, we presented an overview on the existing VITELab infrastructure facilitating remote collaboration by large screen/multi screen visualization. The aim of this infrastructure is to reduce the development costs by simulating and virtually testing possible deployments before the system is actually physically set up.
We have highlighted connected research questions, as well as explained
the VITELab applications in operating, developing and maintaining
industrial automation facilities. The connection to spatial behavioral models and a related type system for the simulation of industrial automation facilities and the connection to visualization capacities was presented in more detail.


\paragraph{Acknowledgements}
We would like to thank staff from RMIT ITS, PropertyServices, eResearch and the VITELab team, in particular Lasith Fernando, Ravi Sreenivasamurthy and Garry Keltie.

\bibliographystyle{apalike}

{\small
\begin{thebibliography}{}

\bibitem[Anderson and Fritzson, 2013]{modelica2013vehicle}
Anderson, A. and Fritzson, P. (2013).
\newblock {Models for Distributed Real-Time Simulation in a Vehicle
  Co-Simulator Setup}.
\newblock In Nilsson, H., editor, {\em {Proceedings of the 5th International
  Workshop on Equation-Based Object-Oriented Modeling Languages and Tools}}.
  Linkoping University Electronic Press.

\bibitem[Basu et~al., 2006]{basu2006modeling}
Basu, A., Bozga, M., and Sifakis, J. (2006).
\newblock Modeling heterogeneous real-time components in bip.
\newblock In {\em 4th IEEE International Conference on Software Engineering and
  Formal Methods (SEFM)}, pages 3--12. IEEE.

\bibitem[Beek et~al., 2006]{Beek06syntaxand}
Beek, D. A.~V., Man, K.~L., Reniers, M.~A., Rooda, J.~E., and Schiffelers, R.
  R.~H. (2006).
\newblock {Syntax and consistent equation semantics of hybrid Chi}.
\newblock In {\em Journal of Logic and Algebraic Programming}, pages 129--210.

\bibitem[Behrmann et~al., 2004]{DBLP:conf/sfm/BehrmannDL04}
Behrmann, G., David, A., and Larsen, K. (2004).
\newblock {A Tutorial on Uppaal}.
\newblock In Bernardo, M. and Corradini, F., editors, {\em Formal Methods for
  the Design of Real-Time Systems}, volume 3185 of {\em LNCS}, pages 200--236.
  Springer.

\bibitem[Bennett et~al., 2002]{bennett}
Bennett, B., Cohn, A.~G., Wolter, F., and Zakharyaschev, M. (2002).
\newblock Multi-dimensional modal logic as a framework for spatio-temporal
  reasoning.
\newblock {\em Applied Intelligence}, 17(3):239--251.

\bibitem[Berger, 2002]{ES_Berger}
Berger, A. (2002).
\newblock {\em Embedded Systems Design: An Introduction to Processes, Tools,
  and Techniques}.
\newblock CMP Books.

\bibitem[Berners-Lee et~al., 2001]{berners2001semantic}
Berners-Lee, T., Hendler, J., Lassila, O., et~al. (2001).
\newblock The semantic web.
\newblock {\em Scientific american}, 284(5):28--37.

\bibitem[Blech, 2013]{DBLP:journals/corr/abs-1302-5175}
Blech, J.~O. (2013).
\newblock Towards a framework for behavioral specifications of osgi components.
\newblock In {\em 11th International Workshop on Formal Engineering approaches
  to Software Components and Architectures (FESCA)}, pages 79--93.

\bibitem[Blech et~al., 2012]{blech2012behavioral}
Blech, J.~O., Falcone, Y., Rue{\ss}, H., and Sch{\"a}tz, B. (2012).
\newblock Behavioral specification based runtime monitors for osgi services.
\newblock In {\em Leveraging Applications of Formal Methods, Verification and
  Validation. Technologies for Mastering Change}, pages 405--419. Springer
  Berlin Heidelberg.

\bibitem[Blech and Sch{\"a}tz, 2012]{blech2012towards}
Blech, J.~O. and Sch{\"a}tz, B. (2012).
\newblock Towards a formal foundation of behavioral types for uml
  state-machines.
\newblock {\em ACM SIGSOFT Software Engineering Notes}, 37(4):1--8.

\bibitem[Blech and Schmidt, 2013]{blechtowards}
Blech, J.~O. and Schmidt, H. (2013).
\newblock Towards modeling and checking the spatial and interaction behavior of
  widely distributed systems.
\newblock In {\em {Improving Systems and Software Engineering Conference}}.

\bibitem[Broy et~al., 2005]{Broy2005testing}
Broy, M., Jonsson, B., Katoen, J.-P., Leucker, M., and Pretschner, A. (2005).
\newblock {\em Model-Based Testing of Reactive Systems: Advanced Lectures
  (LNCS)}.
\newblock Springer.

\bibitem[Cardelli and Gardner, 2010]{cardelli2010processes}
Cardelli, L. and Gardner, P. (2010).
\newblock Processes in space.
\newblock In {\em Programs, Proofs, Processes}, pages 78--87. Springer.

\bibitem[Dal~Zilio et~al., 2004]{zilio}
Dal~Zilio, S., Lugiez, D., and Meyssonnier, C. (2004).
\newblock A logic you can count on.
\newblock In {\em ACM SIGPLAN Notices}, volume~39, pages 135--146. ACM.

\bibitem[de~Alfaro and Henzinger, 2001]{deAlfaro:2001:IA:503271.503226}
de~Alfaro, L. and Henzinger, T.~A. (2001).
\newblock Interface automata.
\newblock {\em SIGSOFT Softw. Eng. Notes}, 26(5):109--120.

\bibitem[De~Moura and Bj{\o}rner, 2008]{de2008z3}
De~Moura, L. and Bj{\o}rner, N. (2008).
\newblock Z3: An efficient smt solver.
\newblock In {\em Tools and Algorithms for the Construction and Analysis of
  Systems}, pages 337--340. Springer.

\bibitem[Deshpande et~al., 1997]{Deshpande97shift:a}
Deshpande, A., Göllü, A., Gollu, A., and Varaiya, P. (1997).
\newblock {Shift: A Formalism and a Programming Language for Dynamic Networks
  of Hybrid Automata}.

\bibitem[Dominka et~al., 2007]{dominka2007hybrid}
Dominka, S., Schiller, F., and Kain, S. (2007).
\newblock Hybrid commissioning—from hardware-in-the-loop simulation to real
  production plants.
\newblock In {\em Proceedings of the 18th IASTED International Conference on
  Modeling and Simulation (MS'07)}, pages 544--549.

\bibitem[Donath et~al., 2008]{modelica2008}
Donath, U., Haufe, J., Blochwitz, T., and Neidhold, T. (2008).
\newblock {A new Approach for Modeling and Verification of Discrete Control
  Components within a Modelica Environment}.

\bibitem[Frehse et~al., 2011]{frehse2011}
Frehse, G., Le~Guernic, C., Donz{\'e}, A., Cotton, S., Ray, R., Lebeltel, O.,
  Ripado, R., Girard, A., Dang, T., and Maler, O. (2011).
\newblock Spaceex: Scalable verification of hybrid systems.
\newblock In {\em Computer Aided Verification}, pages 379--395. Springer.

\bibitem[Fritzson, 2004]{Fritzson2004Modelica}
Fritzson, P. (2004).
\newblock {\em Principles of Object-Oriented Modeling and Simulation with
  Modelica 2.1}.
\newblock Wiley-IEEE Computer Society Press.

\bibitem[Fritzson, 2011]{Fritzson2011Modelica}
Fritzson, P. (2011).
\newblock {\em Introduction to Modeling and Simulation of Technical and
  Physical Systems with Modelica}.
\newblock Wiley-IEEE Computer Society Press.

\bibitem[Hadlich et~al., 2011]{IEEE_INDIN_2011}
Hadlich, T., Diedrich, C., Eckert, K., Frank, T., Fay, A., and Vogel-Heuser, B.
  (2011).
\newblock Common communication model for distributed automation systems.
\newblock In {\em 9th IEEE International Conference on Industrial Informatics},
  IEEE INDIN.

\bibitem[Hazra et~al., 2013]{Daggupta2012tests}
Hazra, A., Ghosh, P., Vadlamudi, S.~G., Chakrabarti, P.~P., and Dasgupta, P.
  (2013).
\newblock Formal methods for early analysis of functional reliability in
  component-based embedded applications.
\newblock {\em Embedded Systems Letters}, 5(1):8--11.

\bibitem[Hirschkoff et~al., 2003]{hirschkoff}
Hirschkoff, D., Lozes, {\'E}., and Sangiorgi, D. (2003).
\newblock Minimality results for the spatial logics.
\newblock In {\em FST TCS 2003: Foundations of Software Technology and
  Theoretical Computer Science}, pages 252--264. Springer.

\bibitem[Isermann et~al., 1999]{isermann1999hardware}
Isermann, R., Schaffnit, J., and Sinsel, S. (1999).
\newblock Hardware-in-the-loop simulation for the design and testing of
  engine-control systems.
\newblock {\em Control Engineering Practice}, 7(5):643--653.

\bibitem[Kratz et~al., 2006]{KratzSPL06}
Kratz, F., Sokolsky, O., Pappas, G.~J., and Lee, I. (2006).
\newblock {R-Charon, a Modeling Language for Reconfigurable Hybrid Systems}.
\newblock In {\em Hybrid Systems: Computation and Control (HSCC)}, pages
  392--406.

\bibitem[Peake et~al., 2013]{peaketowards}
Peake, I., Blech, J.~O., and Fernando, L. (2013).
\newblock Towards reconstructing architectural models of software tools by
  runtime analysis.
\newblock In {\em 3rd International Workshop on Experiences and Empirical
  Studies in Software Modelling}.

\bibitem[Pretschner and Philipps, 2005]{pretschner200510}
Pretschner, A. and Philipps, J. (2005).
\newblock {Methodological Issues in Model-Based Testing}.
\newblock {\em Model-Based Testing of Reactive Systems}, pages 181--291.

\bibitem[Sapienza et~al., 2012]{Sapienza2690}
Sapienza, G., Crnkovic, I., and Seceleanu, T. (2012).
\newblock Towards a methodology for hardware and software design separation in
  embedded systems.
\newblock In {\em Proc. of the Seventh International Conference on Software
  Engineering Advances (ICSEA)}, pages 557--562. IARIA.

\bibitem[Schlager, 2008]{schlager2008hardware}
Schlager, M. (2008).
\newblock Hardware-in-the-loop simulation.

\bibitem[Spichkova, 2007]{spichkova}
Spichkova, M. (2007).
\newblock {\em {Specification and Seamless Verification of Embedded Real-Time
  Systems: FOCUS on Isabelle}}.
\newblock PhD thesis, Technische Universit{\"a}t M{\"u}nchen.

\bibitem[Spichkova, 2012]{hffm_spichkova}
Spichkova, M. (2012).
\newblock {Human Factors of Formal Methods}.
\newblock In {\em {Proc. of IADIS Interfaces and Human Computer Interaction}}.
  IHCI 2012.

\bibitem[Spichkova, 2013a]{Spichkova2013HFFM}
Spichkova, M. (2013a).
\newblock Design of formal languages and interfaces: ``formal'' does not mean
  ``unreadable''.
\newblock In Blashki, K. and Isaias, P., editors, {\em Emerging Research and
  Trends in Interactivity and the Human-Computer Interface}. IGI Global.

\bibitem[Spichkova, 2013b]{IsabelleAFP2013}
Spichkova, M. (2013b).
\newblock {Stream Processing Components: Isabelle/HOL Formalisation and Case
  Studies}.
\newblock {\em {Archive of Formal Proofs}}.

\bibitem[Spichkova and Campetelli, 2012]{Spichkova_Campetelli2012}
Spichkova, M. and Campetelli, A. (2012).
\newblock Towards system development methodologies: From software to
  cyber-physical domain.
\newblock In {\em First International Workshop on Formal Techniques for
  Safety-Critical Systems (FTSCS'12)}.

\bibitem[Spichkova et~al., 2013a]{issec2013spichkova}
Spichkova, M., Schmidt, H., and Peake, I. (2013a).
\newblock {From abstract modelling to remote cyber-physical
  integration/interoperability testing}.
\newblock In {\em {Improving Systems and Software Engineering Conference}}.

\bibitem[Spichkova et~al., 2013b]{spichkova2013we}
Spichkova, M., Zhu, X., and Mou, D. (2013b).
\newblock Do we really need to write documentation for a system?
\newblock In {\em International Conference on Model-Driven Engineering and
  Software Development (MODELSWARD'13)}.

\bibitem[Staab et~al., 2001]{staab2001knowledge}
Staab, S., Studer, R., Schnurr, H.-P., and Sure, Y. (2001).
\newblock Knowledge processes and ontologies.
\newblock {\em Intelligent Systems, IEEE}, 16(1):26--34.

\bibitem[Tundis et~al., 2013]{modelica2013req}
Tundis, A., Rogovchenko-Buffoni, L., Fritzson, P., and Garro, A. (2013).
\newblock {Modeling System Requirements in Modelica: Definition and Comparison
  of Candidate Approaches}.
\newblock In Nilsson, H., editor, {\em {Proceedings of the 5th International
  Workshop on Equation-Based Object-Oriented Modeling Languages and Tools}}.
  Linkoping University Electronic Press.

\bibitem[Vogel-Heuser et~al., 2011]{Vogel-Heuser_IECON}
Vogel-Heuser, B., S., F., Werner, T., and Diedrich, C. (2011).
\newblock Modeling network architecture and time behavior of distributed
  control systems in industrial plant.
\newblock In {\em 37th Annual Conference of the IEEE Industrial Electronics
  Society}, IECON.

\end{thebibliography}
}

\vfill
\end{document}